\documentclass[prd,twocolumn,floatfix,amsmath,nofootinbib,amssymb,floatfix]{revtex4}
\usepackage{graphicx,color,dcolumn,booktabs,bm}
\usepackage{longtable,lscape}
\usepackage{pdfpages}
\usepackage{txfonts}
\usepackage{overpic}
\usepackage{amssymb}
\usepackage{makecell}
\usepackage{indentfirst}
\usepackage{feynmf}   
\usepackage{slashed}  
\usepackage{cases}
\usepackage{color}
\usepackage{multirow}
\usepackage{threeparttable}
\usepackage{enumerate}
\usepackage{subfigure}
\usepackage{diagbox}
\usepackage{siunitx}
\usepackage{mathrsfs}
\usepackage{cancel}
\usepackage{float}
\usepackage[colorlinks,
citecolor=blue,
anchorcolor=red,
menucolor=red,
linkcolor=red,
filecolor=red,
runcolor=red,
urlcolor=blue,
frenchlinks=red]{hyperref}

\usepackage{epstopdf}
\usepackage{amsmath}
\usepackage{graphicx}
\usepackage{adjustbox}
\usepackage{dutchcal}
\usepackage{array}
\usepackage{booktabs}
\usepackage{makecell}
\usepackage{placeins}
\usepackage{textcomp}
\usepackage{amsmath}
\DeclareRobustCommand{\perthousand}{%
	\ifmmode
	\text{\textperthousand}%
	\else
	\textperthousand
	\fi}
\begin{document}
	\title{Analysis of the hadronic molecules $DK$, $D^*K$, $DK^*$ and their bottom analogs with QCD sum rules}
	\author{Ze Zhou$^{1}$}
	\author{Guo-Liang Yu$^{1,2}$}
	\email{yuguoliang2011@163.com}
	\author{Zhi-Gang Wang$^{1,2}$}
	\email{zgwang@aliyun.com}
	\author{Jie Lu$^{3}$}
	\author{Bin Wu$^{1}$}

	\affiliation{$^1$ Department of Mathematics and Physics, North China
		Electric Power University, Baoding 071003, People's Republic of
		China\\$^2$ Hebei Key Laboratory of Physics and Energy Technology, North China Electric Power University, Baoding 071000, China \\$^3$ School of Physics, Southeast University, Nanjing 210094, China}
	\date{\today}
	
\begin{abstract}
In this work, we construct the color-singlet-color-singlet type currents to study the masses and pole residues of charm-strange tetraquark states and their bottom analogs with $J^P$ = $0^+$ and $1^+$ by using two-point QCD sum rules, where the vacuum condensates are considered up to dimension 12. The predicted masses for $DK$, $D^*K$ and $DK^*$ molecular states are $2.322_{ - 0.072}^{ + 0.066}$ GeV, $2.457_{ - 0.068}^{ + 0.064}$ GeV and $2.538_{ - 0.062}^{ + 0.059}$ GeV. These results are consistent well with the experimental data of $D_{s0}(2317)$, $D_{s1}(2460)$ and ${D}_{s1}(2536)$, respectively. The theoretical results for $BK$ and $B^*K$ molecular states are $5.970_{ - 0.064}^{ + 0.061}$ GeV and $6.050_{ - 0.064}^{ + 0.062}$ GeV which are all higher than their own thresholds. Finally, the mass of hadronic molecule $BK^*$ is predicted to be $6.158_{ - 0.063}^{ + 0.061}$ GeV. This value is lower than the threshold of $BK^*$, which implies that it may be a bound hadronic molecular state.
\end{abstract}
\maketitle
	
	\section{Introduction}\label{sec1}
	
	$D_{s0}^*(2317)$ was discovered by BaBar collaboration \cite{BaBar:2003oey} in the mass distribution of $D_s^ + {\pi ^0}$ in the annihilation process of ${e^ + }{e^ - }$ in 2003. In the same year, $D_{s1}(2460)$ treated as the spin partner of $D_{s0}^*(2317)$ in the heavy quark spin symmetry was observed by CLEO collaboration \cite{CLEO:2003ggt}. Subsequently, several other experiments \cite{Belle:2003guh, BaBar:2004yux,BaBar:2006eep} confirmed these discoveries. In addition, another charm-strange state ${D}_{s1}(2536)$ was also measured earlier by Belle collaboration \cite{ARGUS:1989zue}. Since the observation of $D_{s0}^*(2317)$, $D_{s1}(2460)$ and ${D}_{s1}(2536)$, extensive experimental investigations about these states were conducted \cite{Belle:2003kup,BaBar:2004gtd, BaBar:2003cdx, Akhmedov:2005np}. The masses and decay widths of these charm-strange states are listed in Tab. \ref{table for mass and decay width}.
	
	\begin{table}[h!]
		\begin{ruledtabular}
			\centering
			\renewcommand{\arraystretch}{1.4}
			\caption{The experimental data for $D_{s0}^{*}(2317)$, $D_{s1}(2460)$ and $D_{s1}(2536)$ \cite{ParticleDataGroup:2024cfk}.}
			\begin{tabular}{c c c c}
			States & $J^P$ & Mass (MeV) & Full width (MeV)\\
			\hline
			$D_{s0}^{*}(2317)$ & $0^+$ & 2317.8 $\pm$ 0.5 & $\Gamma$ $<$ 3.8\\
			$D_{s1}(2460)$ & $1^+$ & 2459.5 $\pm$ 0.6 & $\Gamma$ $<$ 3.5\\
			$D_{s1}(2536)$ & $1^+$ & 2535.11 $\pm$ 0.06 & $\Gamma$ $=$ 0.92 $\pm$ 0.05
			\end{tabular}
			\label{table for mass and decay width}
		\end{ruledtabular}
	\end{table}

At the same time, there were also a lot of intense theoretical analyses about the underlying structure of these states. In particularly, the masses of $D_{s0}^*(2317)$ and $D_{s1}(2460)$ are about one hundred MeV lower than the predictions of quark model \cite{Godfrey:1985xj, Godfrey:1986wj, DiPierro:2001dwf}. Furthermore, this large deviation also appeared in the simulations of lattice QCD \cite{Moir:2013ub} and greatly confused the theoretical and experimental physicists. This issue is commonly referred to as low-mass issue, and similar phenomenon was also observed in ${\Lambda _c}(2940)$ \cite{Luo:2019qkm} and $X(3872)$ \cite{Kalashnikova:2005ui}. Up to now, there have been many theoretical studies which were proposed to solve this problem and understand the inner structure of these states. For example, $D_{s0}^*(2317)$, $D_{s1}(2460)$ were analyzed as compact tetraquark states \cite{Nielsen:2005ia, Wang:2006uba, Chen:2004dy, Dmitrasinovic:2005gc, Cheng:2003kg, Terasaki:2003qa, Dmitrasinovic:2004cu, Kim:2005gt, Zhang:2018mnm}, or the conventional quark-antiquark states \cite{Bardeen:2003kt, Deandrea:2003gb, Dai:2003yg, Sadzikowski:2003jy, Cahn:2003cw, Hwang:2004cd, Simonov:2004ar, Cheng:2014bca, Song:2015nia, Cheng:2017oqh, Luo:2021dvj, Zhou:2020moj, Alhakami:2016zqx}. In addition, there are also some papers advocating that these states can be explained as a mixture state of quark-antiquark plus four-quark components \cite{Vijande:2006hj, Maiani:2004vq, Dai:2006uz, Browder:2003fk}, or be interpreted as molecular states \cite{vanBeveren:2003kd, Liu:2012zya, Barnes:2003dj, Roca:2025lij, Kong:2021ohg, Kolomeitsev:2003ac, Szczepaniak:2003vy, Hofmann:2003je, Gamermann:2006nm, Flynn:2007ki, Faessler:2007gv, Guo:2009ct, Xie:2010zza, Cleven:2010aw, Wu:2011yb, Guo:2015dha, Albaladejo:2016hae, Du:2017ttu, Guo:2018tjx, Albaladejo:2018mhb, Wu:2019vsy, Gregory:2021rgy, Wang:2012bu, Huang:2021fdt,Chen:2016spr,Liu:2019zoy}. As for ${D}_{s1}(2536)$, there are also many articles that treat it as $DK^*$ tetraquark molecular state \cite{Lin:2024hys,Gamermann:2007fi,Guo:2006fu,Guo:2006rp}, because it could be obtained from the interaction of coupling channel in the molecular picture.
	
As the partner of open charm molecular states, bottom-strange states with $I{(J^P)}$ = $0{(0^+) }$ and $0{(1^+) }$ also attract our attention. Despite the lack of experimental data, theoretical physicists have conducted extensive researches on these states \cite{Guo:2006fu, Guo:2006rp, Faessler:2008vc}. The results indicate that there may exist two bound states $BK$ and $B^*K$ whose energies are below their own thresholds, respectively. Recently, LHCb collaboration \cite{LHCb:2020pet} observed two structures which can be treated as ${B_{sJ}}(6064)$ and ${B_{sJ}}(6114)$ if they decay directly to $B^+K^-$. These two bottom states can also be treated as ${B_{sJ}}(6109)$ and ${B_{sJ}}(6158)$ if decaying through ${B^{* \pm }}{K^ \mp }$ channel. The mass of ${B_{sJ}}(6158)$ is measured to be $6158\pm4\pm5$ MeV which is very close to the threshold of $BK^*$. Thus, the investigation of $BK$, $B^*K$ and $BK^*$ molecular state is an interesting work.
	
That is to say, although there have been many studies on the inner structures of $D_{s0}(2317)$, $D_{s1}(2460)$, ${D}_{s1}(2536)$ and their bottom analogs by using different methods, the conclusions were not consistent well with each other. Thus, it is still very essential for us to study these open charm/bottom states with QCD sum rules \cite{Shifman:1978by, Shifman:1978bx}. As a very powerful no-perturbative approach, the QCD sum rules have been widely used to study the mass spectrum, decay constant, form factor and coupling constant of hadrons \cite{Wei:2006wa, Wang:2024ciy, Wang:2015epa, Wang:2013vex, Aliev:2012tt, Belyaev:1993wp, Dai:1996xv}. The purpose of the present work is to provide valuable information to understand the structures of $D_{s0}(2317)$, $D_{s1}(2460)$, ${D}_{s1}(2536)$ and their bottom analogs by analyzing the hadronic molecular states $DK$, $D^*K$, $DK^*$, $BK$, $B^*K$ and $BK^*$.
	
	This article is arranged as follows: After the introduction, we obtain the QCD sum rules of masses and pole residues of $DK$, $D^*K$, $DK^*$, $BK$, $B^*K$ and $BK^*$ molecular states in section \ref{sec2}. The numerical results and discussions are given in section  \ref{sec3}, Section \ref{sec4} is reserved for a short conclusion.
	
	\section{The two-point QCD sum rule}\label{sec2}
	
	In order to obtain the masses and pole residues of charm-strange tetraquark molecular states $DK$, $D^*K$, $DK^*$ and their bottom analogs, we write down the following two-point correlation functions,
\begin{align}\label{two-point correlatar}
\notag
\Pi^{0} (p) &= i\int {{d^4}x} {e^{ip \cdot x}}\left\langle 0 \right|T\{ J^{0}(x){J^ {0\dag} }(0)\} \left| 0 \right\rangle \\
\notag
		\Pi^{1} _{\mu \nu } (p) &= i\int {{d^4}x} {e^{ip \cdot x}}\left\langle 0 \right|T\{ J^{1}_\mu  (x)J^{1\dag}_{\nu} (0)\} \left| 0 \right\rangle  \\
\Pi^{2} _{\mu \nu } (p) &= i\int {{d^4}x} {e^{ip \cdot x}}\left\langle 0 \right|T\{ J^{2}_\mu  (x)J^{2\dag}_{\nu} (0)\} \left| 0 \right\rangle
\end{align}
	where $T$ is the time ordered product, and $J^{0}(x)$, $J^{1}_{\mu}(x)$ and $J^{2}_{\mu}(x)$ are the interpolating currents with the same quantum numbers as studied states. The currents of these states can be expressed as,
	\begin{align}\label{current of four quark states}
		\notag
		J^{0}(x) =& \frac{1}{{\sqrt 2 }}[{{\bar s}^m}(x)i{\gamma _5}{u^m}(x){{\bar u}^n}(x)i{\gamma _5}{Q^n}(x)\\
		\notag
		& + {{\bar s}^m}(x)i{\gamma _5}{d^m}(x){{\bar d}^n}(x)i{\gamma _5}{Q^n}(x)]\\
		\notag
		J_\mu ^1(x) =& \frac{1}{{\sqrt 2 }}[{{\bar s}^m}(x)i{\gamma _5}{u^m}(x){{\bar u}^n}(x){\gamma _\mu }{Q^n}(x)\\
		\notag
		& + {{\bar s}^m}(x)i{\gamma _5}{d^m}(x){{\bar d}^n}(x){\gamma _\mu }{Q^n}(x)]\\
		\notag
		J_\mu ^2(x) =& \frac{1}{{\sqrt 2 }}[{{\bar s}^m}(x){\gamma _\mu }{u^m}(x){{\bar u}^n}(x)i{\gamma _5}{Q^n}(x)\\
		& + {{\bar s}^m}(x){\gamma _\mu }{d^m}(x){{\bar d}^n}(x)i{\gamma _5}{Q^n}(x)]
	\end{align}
Here, $m$, $n$ are color indexes, and $Q$ stands for $c$ or $b$ quark field. The currents $J^{0}(x)$, $J_\mu ^1(x)$ and $J_\mu ^2(x)$ represent $DK$, $D^*K$ and $DK^*$ tetraquark molecular states with $Q=c$ or their bottom analogs with $Q=b$.
	
In the framework of QCD sum rules, the correlation function will be dealt at both phenomenological side and QCD side. Then, according quark-hadron duality, we can obtain the sum rules for masses and pole residues of tetraquark molecular states in the hadronic level.
	
	\subsection{The phenomenological side}
	In the phenomenological side, a complete set of intermediate hadronic states with the same quantum numbers as the current operators are inserted into correlation function to obtain the phenomenological representation,
	\begin{align}
		\notag
		\Pi^{\texttt{phy}0} (p) &= \frac{{\left\langle 0 \right|J(0)\left| {S(p)} \right\rangle \left\langle {S(p)} \right|{J^ \dag }(0)\left| 0 \right\rangle }}{{m_S^2 - {p^2}}} +  \cdot  \cdot  \cdot \\
		\notag
		\Pi^{\texttt{phy}1(2)}_{\mu \nu }(p) &= \frac{{\left\langle 0 \right|J_\mu ^{1(2)}(0)\left| {A_{1(2)}(p)} \right\rangle \left\langle {A_{1(2)}(p)} \right|J{{_\nu ^{1(2)}}^ \dag }(0)\left| 0 \right\rangle }}{{m_{{A_{1(2)}}}^2 - {p^2}}}\\
		&+ \frac{{\left\langle 0 \right|J_\mu ^{1(2)}(0)\left| {P_{1(2)}(p)} \right\rangle \left\langle {P_{1(2)}(p)} \right|J{{_\nu ^{1(2)}}^ \dag }(0)\left| 0 \right\rangle }}{{m_{P_{1(2)}}^2 - {p^2}}} +  \cdot  \cdot  \cdot
	\end{align}
Here, $S$, $A_{1(2)}$ and $P_{1(2)}$ stand for scalar ($DK$/$BK$), axial-vector ($D^{*}K$/$B^{*}K$ and $DK^{*}$/$BK^{*}$) and pseudoscalar particles which can couple with currents in Eq. (\ref{current of four quark states}), respectively, and ellipsis denote the contributions from excited and continuum states. The above vacuum matrix elements can be defined as follows,	
	\begin{align}\label{definite matrix of mass}
		\notag
		\left\langle 0 \right|J^{0}(0)\left| {S(p)} \right\rangle  &= {f_S} \\
		\notag
		\left\langle 0 \right|{J_\mu^{1(2)} }(0)\left| {P_{1(2)}(p)} \right\rangle  &= \frac{{f_{P_{1(2)}}{p_\mu }}}{{m_{P_{1(2)}}}} \\
		\left\langle 0 \right|J_\mu ^{1(2)}(0)\left| {{A_{1(2)}(p)}} \right\rangle  &= {f_{A_{1(2)}}}{\varepsilon _\mu }
	\end{align}
where $f_{S}$, $f_{P_{1(2)}}$ and $f_{A_{1(2)}}$ represent the pole residues, and ${\varepsilon _\mu }$ is the polarization vector of axialvector tetraquark molecular state. According to dispersion relation, we obtain the following representation at the phenomenological side where the contributions of excited and continuum states have been omitted,
		\begin{align}\label{hadronic expression}
		\notag
		\Pi^{\texttt{phy}0} ({p}) &= \frac{{f_S^2}}{{m_S^2 - {p^2}}}  = {\Pi ^S}({p^2})\\
		\notag
		{\Pi _{\mu \nu }^{\texttt{phy}1}}({p}) &= \frac{{f^{2}_{A_{1}}}}{{m_{A_{1}}^{2} - {p^2}}}{{\tilde g}_{\mu \nu }} + \frac{f_{P_{1}}^{2}}{{m^{2}_{P_{1}} - {p^2}}}{p_\mu }{p_\nu } \\
		\notag
		&= \Pi _{{g_{\mu \nu }}}^{A_{1}}({p^2}){g_{\mu \nu }} + \Pi^{P_{1}} _{{p_\mu }{p_\nu }}({p^2}){p_\mu }{p_\nu }\\
		\notag
		\Pi _{\mu \nu }^{\texttt{phy}2}({p}) &= \frac{{f_{A_{2}}^2}}{{m_{A_{2}}^2 - {p^2}}}{{\tilde g}_{\mu \nu }} + \frac{{f_{P_{2}}^2}}{{m_{P_{2}}^2 - {p^2}}}{p_\mu }{p_\nu } \\
		&= \Pi _{{g_{\mu \nu }}}^{A_{2}}({p^2}){g_{\mu \nu }} + \Pi^{P_{2}} _{{p_\mu }{p_\nu }}({p^2}){p_\mu }{p_\nu }
	\end{align}
	where ${\Pi^S}({p^2})$, $\Pi _{{g_{\mu \nu }}}^{A_{1(2)}}({p^2})$ and $\Pi^{P_{1(2)}} _{{p_\mu }{p_\nu }}({p^2})$ are the scalar invariant amplitudes associated with different Lorentz structures, and ${{\tilde g}_{\mu \nu }}$ has the following property,
	\begin{align}\label{polerazation sum}
		{{\tilde g}_{\mu \nu }} =\mathop \sum \limits_\lambda  \varepsilon _\mu ^*(\lambda ,p){\varepsilon _\nu }(\lambda ,p)=  - {g_{\mu \nu }} + \frac{{{p_\mu }{p_\nu }}}{{{p^2}}},
	\end{align}

\subsection{The QCD side}
After performing wick's contraction, the correlation function can be written as the following forms in the QCD side,
\begin{align}\label{QCD contract Pi(x)}
	\notag
	{\Pi ^{\mathrm{QCD}0}}(p) =& i\int {{d^4}x} {e^{ip \cdot x}}Tr[{U^{n'n}}( - x){\gamma _5}{Q^{nn'}}(x){\gamma _5}] \\
	\notag
	& \times Tr[{U^{mm'}}(x){\gamma _5}{S^{m'm}}( - x){\gamma _5}],\\
	\notag
	\Pi _{\mu \nu }^{\mathrm{QCD1}}(p) =&  - i\int {{d^4}x} {e^{ip \cdot x}}Tr[{U^{n'n}}( - x){\gamma _\mu }{Q^{nn'}}(x){\gamma _\nu }]\\
	\notag
	& \times Tr[{U^{mm'}}(x){\gamma _5}{S^{m'm}}( - x){\gamma _5}]\\
	\notag
	\Pi _{\mu \nu }^{\mathrm{QCD2}}(p) =&  - i\int {{d^4}x} {e^{ip \cdot x}}Tr[{U^{n'n}}( - x){\gamma _5}{Q^{nn'}}(x){\gamma _5}]\\
	& \times Tr[{U^{mm'}}(x){\gamma _\nu }{S^{m'm}}( - x){\gamma _\mu }],
\end{align}
where $U^{ij}$ , $S^{ij}$ and $Q^{ij}$ are the $u/d$, $s$, and $c/b$ quark propagators and have the following expressions,

\begin{align}\label{light quark propagator}
	\notag
	{q^{ij}}(x) = &\frac{{i{\delta ^{ij}}{\slashed{x}}}}{{2{\pi ^2}{x^4}}} - \frac{{{\delta ^{ij}}{m_q}}}{{4{\pi ^2}{x^4}}} - \frac{{{\delta ^{ij}}\left\langle {\bar qq} \right\rangle }}{{12}} + \frac{{i{\delta ^{ij}}\slashed{x}{m_q}\left\langle {\bar qq} \right\rangle }}{{48}}\\
	\notag
	&- \frac{{{\delta ^{ij}}{x^2}\left\langle {\bar q{g_s}\sigma Gq} \right\rangle }}{{192}} + \frac{{i{\delta ^{ij}}{x^2}\slashed{x}{m_q}\left\langle {\bar q{g_s}\sigma Gq} \right\rangle }}{{1152}}\\
	\notag
	&- \frac{{i{g_s}G_{\alpha \beta }^at_{ij}^a(\slashed{x}{\sigma ^{\alpha \beta }} + {\sigma ^{\alpha \beta }}\slashed{x})}}{{32{\pi ^2}{x^2}}} - \frac{{i{\delta ^{ij}}{x^2}\slashed{x}g_s^2{{\left\langle {\bar qq} \right\rangle }^2}}}{{7776}}\\
	\notag
	&- \frac{{{\delta ^{ij}}{x^4}\left\langle {\bar qq} \right\rangle \left\langle {g_s^2GG} \right\rangle }}{{27648}} - \frac{{\left\langle {{{\bar q}^i}{\sigma ^{\mu \nu }}{q^j}} \right\rangle {\sigma _{\mu \nu }}}}{8} - \\
	&\frac{{\left\langle {{{\bar q}^i}{\gamma ^\mu }{q^j}} \right\rangle {\gamma _\mu }}}{4}\cdots
\end{align}
\begin{align}\label{heavy quark propagator}
	\notag
	{Q^{ij}}(x) = &\frac{i}{{{{(2\pi )}^4}}}\int {{d^4}k} {e^{ - ik \cdot x}}\Big\{ \frac{{{\delta ^{ij}}(\slashed{k} + {m_Q})}}{{({k^2} - m_Q^2)}}\\
	\notag
	&- \frac{{{g_s}G_{\alpha \beta }^nt_{ij}^n}}{4}\frac{{{\sigma ^{\alpha \beta }}(\slashed{k} + {m_Q}) + (\slashed{k} + {m_Q}){\sigma ^{\alpha \beta }}}}{{{{({k^2} - m_Q^2)}^2}}}\\
	\notag
	&+ \frac{{{g_s}{D_\alpha }G_{\beta \lambda }^nt_{ij}^n({f^{\lambda \beta \alpha }} + {f^{\lambda \alpha \beta }})}}{{3{{({k^2} - m_Q^2)}^4}}}\\
	&- \frac{{g_s^2{{({t^a}{t^b})}_{ij}}G_{\alpha \beta }^aG_{\mu \nu }^b({f^{\alpha \beta \mu \nu }} + {f^{\alpha \mu \beta \nu }} + {f^{\alpha \mu \nu \beta }})}}{{4{{({k^2} - m_Q^2)}^5}}} \cdots  \Big\}
\end{align}
Here, $q^{ij}$ represents $U^{ij}$ and $S^{ij}$ quark propagators, the superscript $i$ and $j$ are color indexes, and ${t^n} = {\lambda ^n}/2$. ${\lambda^n}$ is the Gell-Mann matrix, ${\sigma _{\alpha \beta }} = i[{\gamma _\alpha },{\gamma _\beta }]/2$, ${D_\alpha } = {\partial _\alpha } - i{g_s}G_\alpha ^n{t^n}$ with $G_\alpha ^n$ being the gluon field. In the full propagator of light quark, the term $\left\langle {{{\bar q}^i}{\sigma ^{\mu \nu }}{q^j}} \right\rangle $ which comes from Fierz re-arrangement is reserved to absorb the gluons emitted from the other quark lines. In the full propagator of heavy quark, ${f^{\lambda \alpha \beta  }}$ and ${f^{\alpha \beta \mu \nu }}$ have the following expressions,

\begin{align}\label{f of double GG definition}
	\notag
	{f^{\lambda \alpha \beta  }} =& (\slashed{k} + {m_Q}){\gamma ^\lambda }(\slashed{k} + {m_Q}){\gamma ^\alpha }(\slashed{k} + {m_Q}){\gamma ^\beta }(\slashed{k} + {m_Q})\\
	\notag
	{f^{\alpha \beta \mu \nu }} = &(\slashed{k} + {m_Q}){\gamma ^\alpha }(\slashed{k} + {m_Q}){\gamma ^\beta }(\slashed{k} + {m_Q})\\
	&\times {\gamma ^\mu }(\slashed{k} + {m_Q}){\gamma ^\nu }(\slashed{k} + {m_Q})
\end{align}

The above full propagators are put into Eq. (\ref{QCD contract Pi(x)}) to further calculate the correlation function, and the correlation function in the QCD side can also be expanded in different tensor structures which are the same as Eq. (\ref{hadronic expression}) in the phenomenological side,
\begin{align}
	\notag
	\Pi^{\mathrm{QCD}0} (p) &= {{\Pi} ^{\mathrm{QCD}0}}({p^2})\\
	\notag
	{\Pi _{\mu \nu }^{\mathrm{QCD1}}}(p) &= \Pi _{{g_{\mu \nu }}}^{\mathrm{QCD1}}({p^2}){g_{\mu \nu }} + \Pi _{{p_\mu }{p_\nu }}^{\mathrm{QCD1}}({p^2}){p_\mu }{p_\nu }\\
	{\Pi _{\mu \nu }^{\mathrm{QCD2}}}(p) &= \Pi _{{g_{\mu \nu }}}^{\mathrm{QCD2}}({p^2}){g_{\mu \nu }} + \Pi _{{p_\mu }{p_\nu }}^{\mathrm{QCD2}}({p^2}){p_\mu }{p_\nu }
\end{align}
where $\Pi^{\mathrm{QCD}0}(p^2)$, $\Pi^{\mathrm{QCD1(2)}}_{g_{\mu \nu }}(p^2)$ and $\Pi^{\mathrm{QCD1(2)}}_{{p_\mu }{p_\nu }}(p^2)$ are scalar invariant amplitude. According to the dispersion relations, the correlation function can be written as the following expression,
\begin{align}\label{dis rela}
	\Pi ^{\mathrm{QCD}}(p^2) = \int\limits_{{s_{\min }}}^{{s_0}} {ds} \frac{{\rho ^{\mathrm{QCD}}(s)}}{{s - {p^2}}}
\end{align}
where $\rho^{QCD}(s)$ is the QCD spectral density and it is obtained by taking the imaginary part of the correlation functions. In order to extract the pure contributions from different structures, we choose scalar invariant amplitudes ${{\Pi} ^{\mathrm{QCD0}}}({p^2})$, $\Pi _{{g_{\mu \nu }}}^{\mathrm{QCD1}}({p^2})$ and $\Pi _{{g_{\mu \nu }}}^{\mathrm{QCD2}}({p^2})$ to study the scalar and axial-vector tetraquark molecular states. For $\Pi_{\mu\nu}^{QCD1}$ as an example, its spectral density is obtained by the following relation,
\begin{align}
\rho^{\mathrm{QCD}1}_{\mu\nu}(s)=\frac{1}{\pi}\texttt{Im}[\Pi^{\mathrm{QCD}}_{g_{\mu\nu}}(s+i\varepsilon)]
\end{align}
For simplicity, only the QCD spectral density of the $DK/BK$ molecular state $\rho^{QCD0}(s)$ is listed in Appendix \ref{appendix B}. In Eq. (\ref{dis rela}), $s_{min}$ is kinematic limit, $s_0$ is continuum threshold parameter and it can be presented as ${(M + \Delta )^2}$, where $M$ is the mass of the ground state and $\Delta $ is the energy gap between ground state and the first excited state. Its value is often taken as $\Delta=$0.3 GeV$-$0.7 GeV.

After matching the phenomenological side and QCD side of correlation function and performing Borel transformation with regard to the $P^2=-p^2$, the following QCD sum rules are acquired,
\begin{align}\label{decay constant sum rule}
	\notag
	f_S^2\exp \left( { -\frac{{m_S^2}}{{{T^2}}}} \right) &= \int\limits_{m_Q^2}^{{s_0}} {ds} {\rho ^{\mathrm{QCD}0}(s)}\exp \left( { - \frac{s}{{{T^2}}}} \right) \\
	- f_{A_{1(2)}}^2\exp \left( { - \frac{{m_{A_{1(2)}}^2}}{{{T^2}}}} \right) &= \int\limits_{m_Q^2}^{{s_0}} {ds} {\rho _{{{g_{\mu \nu }}}}^{\mathrm{QCD}1(2)}(s)}\exp \left( { - \frac{s}{{{T^2}}}} \right)
\end{align}
where $T^2$ is the Borel parameter.

Finally, we perform differentiation with respect to the variable $\tau  = \frac{1}{{{T^2}}}$ in Eq. (\ref{decay constant sum rule}), then, the QCD sum rules for the mass of tetraquark molecular states are obtained,
\begin{align}\label{mass sum rule}
\notag
m_{S}^2 =  \frac{{\int\limits_{m_Q^2}^{{s_0}} {ds} s{\rho _{{{g_{\mu \nu }}}}^{\mathrm{QCD}0}}(s)\exp \left( { - \tau s} \right)}}{{\int\limits_{m_Q^2}^{{s_0}} {ds} {\rho ^{\mathrm{QCD}0}}(s)\exp \left( { - \tau s} \right)}} \\
m_{A_{1(2)}}^2 =  \frac{{\int\limits_{m_Q^2}^{{s_0}} {ds} s{\rho _{{{g_{\mu \nu }}}}^{\mathrm{QCD}1(2)}}(s)\exp \left( { - \tau s} \right)}}{{\int\limits_{m_Q^2}^{{s_0}} {ds} {\rho _{{{g_{\mu \nu }}}}^{\mathrm{QCD}1(2)}}(s)\exp \left( { - \tau s} \right)}}
\end{align}

\section{Numerical result for masses and pole residues}\label{sec3}

The final numerical results are dependent on some input parameters such as the mass of heavy quark and vacuum condensates. The standard values of vacuum condensates are taken as $\left\langle {\bar qq} \right\rangle =- {(0.24 \pm 0.01 \ \mathrm{GeV})^3},$ $\left\langle {\bar ss} \right\rangle=(0.8 \pm 0.1)\left\langle {\bar qq} \right\rangle ,$ $\left\langle {\bar q{g_s}\sigma Gq} \right\rangle=m_0^2\left\langle{\bar qq} \right\rangle,$ $\left\langle {\bar s{g_s}\sigma Gs} \right\rangle=m_0^2\left\langle {\bar ss} \right\rangle,$ $m_0^2=(0.8 \pm 0.1)\ \mathrm{GeV^2},$ $\left\langle{\frac{{{\alpha_s}GG}}{\pi }}\right\rangle=(0.012 \pm 0.004) \ \mathrm{GeV^4}$ at the energy scalar $\mu$=1 GeV with $q=u$ and $d$ quarks \cite{Shifman:1978by, Shifman:1978bx, Reinders:1984sr}.

It is noted that the the masses of $c$ and $b$ quarks and the values of vacuum condensates are energy dependent, which can be expressed as follows by using the renormlization group equation(RGE),
\begin{align}\label{vacuum codensation}
	\notag
	\left\langle {\bar qq} \right\rangle (\mu ) &= \left\langle {\bar qq} \right\rangle (\mathrm{1GeV}){\left[ {\frac{{\alpha _s}(\mathrm{1GeV})}{{\alpha _s}(\mu )}} \right]^{\frac{{12}}{{33 - 2{n_f}}}}} \\
		\notag
	\left\langle {\bar q{g_s}\sigma Gq} \right\rangle (\mu ) &= \left\langle {\bar q{g_s}\sigma Gq} \right\rangle (\mathrm{1GeV}){\left[ {\frac{{\alpha _s}(\mathrm{1GeV})}{{\alpha _s}(\mu )}} \right]^{\frac{2}{{33 - 2{n_f}}}}} \\
	\notag
	{m_Q}(\mu ) &= {m_Q}({m_Q}){\left[ {\frac{{{\alpha _s}(\mu )}}{{{\alpha _s}({m_Q})}}} \right]^{\frac{{12}}{{33 - 2{n_f}}}}} \\
	\notag
	{m_s}(\mu ) &= {m_s}(2 \mathrm{GeV}){\left[ {\frac{{{\alpha _s}(\mu )}}{{{\alpha _s}(2\mathrm{GeV})}}} \right]^{\frac{{12}}{{33 - 2{n_f}}}}} \\
	{\alpha _s}(\mu ) = &\frac{1}{{{b_0}t}}\left[ {1 - \frac{{{b_1}}}{{b_0^2}}\frac{{\log t}}{t} + \frac{{b_1^2({{\log }^2}t - \log t - 1) + {b_0}{b_2}}}{{b_0^4{t^2}}}} \right]
\end{align}
where $q$ = $u$, $d$ or $s$ quark, $Q$ denote heavy quark $c$ or $b$, $t = \log \frac{{{\mu ^2}}}{{\Lambda _{\mathrm{QCD}}^2}},$ ${b_0} = \frac{{33 - 2{n_f}}}{{12\pi }},$ ${b_1} = \frac{{153 - 19{n_f}}}{{24{\pi ^2}}},$ ${b_2} = \frac{{2857 - \frac{{5033}}{9}{n_f} + \frac{{325}}{{27}}n_f^2}}{{128{\pi ^3}}}. $ ${\Lambda _{\mathrm{QCD}}}$ = 210 MeV, 292MeV and 332MeV for the flavors $n_f$ = 5, 4 and 3, respectively, and $n_f$ is flavor number \cite{ParticleDataGroup:2024cfk}. In this work, we set $n_f$ = 4 and $n_f$ = 5 for the charm-strange and bottom-strange tetraquark molecular states. Thus, it is crucial to choose a suitable energy scale to determine the masses and pole residues of tetraquark molecular states. In our previous work, a energy scale formula $\mu =\sqrt {M^2 - \mathbb{M}_{c/b}^2}-k{\mathbb{M}_s}$ is developed to determine the optimal energy scale \cite{Wang:2013daa,Wang:2025sic}, where $M$ denotes the mass of tetraquark molecular state, $k$ is the number of $s$ quark in the currents, and $\mathbb{M}_{c/b}$, $\mathbb{M}_{s}$ are the effective masses of $c/b$ and $s$ quarks. In the present work, the values $\mathbb{M}_{c}=1.82$ GeV, $\mathbb{M}_{b}=5.17$ GeV and $\mathbb{M}_{s}=0.20$ GeV are chosen for the tetraquark molecular states. The modified-minimal-subtraction masses are adopted as ${m_c}({m_c})=(1.275 \pm 0.025)$ GeV, ${m_b}({m_b})=(4.18 \pm 0.03)\ \mathrm{GeV}$ and ${m_s}(2 \mathrm{GeV})=(0.095 \pm 0.005) \ \mathrm{GeV}$ from the Particle Data Group \cite{ParticleDataGroup:2024cfk}.

It can be seen from Eq. (\ref{mass sum rule}) that the final results depend also on Borel parameter $T^2$ and continuum threshold $s_0$.  To obtain reliable results, an appropriate working region of Borel parameter should be selected. The hadron parameters like mass and pole residue should have a weak Borel parameter dependency in this region. This working region is commonly named as Borel platform or Borel window. At the same time, three criteria should be satisfied, which are pole dominance, convergence of operator product expansion (OPE) and satisfying the energy scale formula. The pole contribution is defined as,
\begin{align}\label{PC}
	\mathrm{PC} = \frac{{\int\limits_{m_Q^2}^{{s_0}} {ds} {\rho ^{\mathrm{QCD}}}(s)\exp \left( { - \frac{s}{{{T^2}}}} \right)}}{{\int\limits_{m_Q^2}^\infty  {ds} {\rho ^{\mathrm{QCD}}}(s)\exp \left( { - \frac{s}{{{T^2}}}} \right)}}
\end{align}
The contribution of vacuum condensate of dimension $n$ is defined as,
\begin{align}\label{condensation of vacuum}
	\mathrm{D(n)} = \frac{{\int\limits_{m_Q^2}^{{s_0}} {ds} {\rho _n^{\mathrm{QCD}}}(s)\exp \left( { - \frac{s}{{{T^2}}}} \right)}}{{\int\limits_{m_Q^2}^{{s_0}} {ds} {\rho ^{\mathrm{QCD}}}(s)\exp \left( { - \frac{s}{{{T^2}}}} \right)}}
\end{align}
where ${\rho _n^{\mathrm{QCD}}}$ denotes the spectral density of dimension $n$.

\begin{figure}
	\centering
	\includegraphics[width=8.5cm, trim=0 0 20 30, clip]{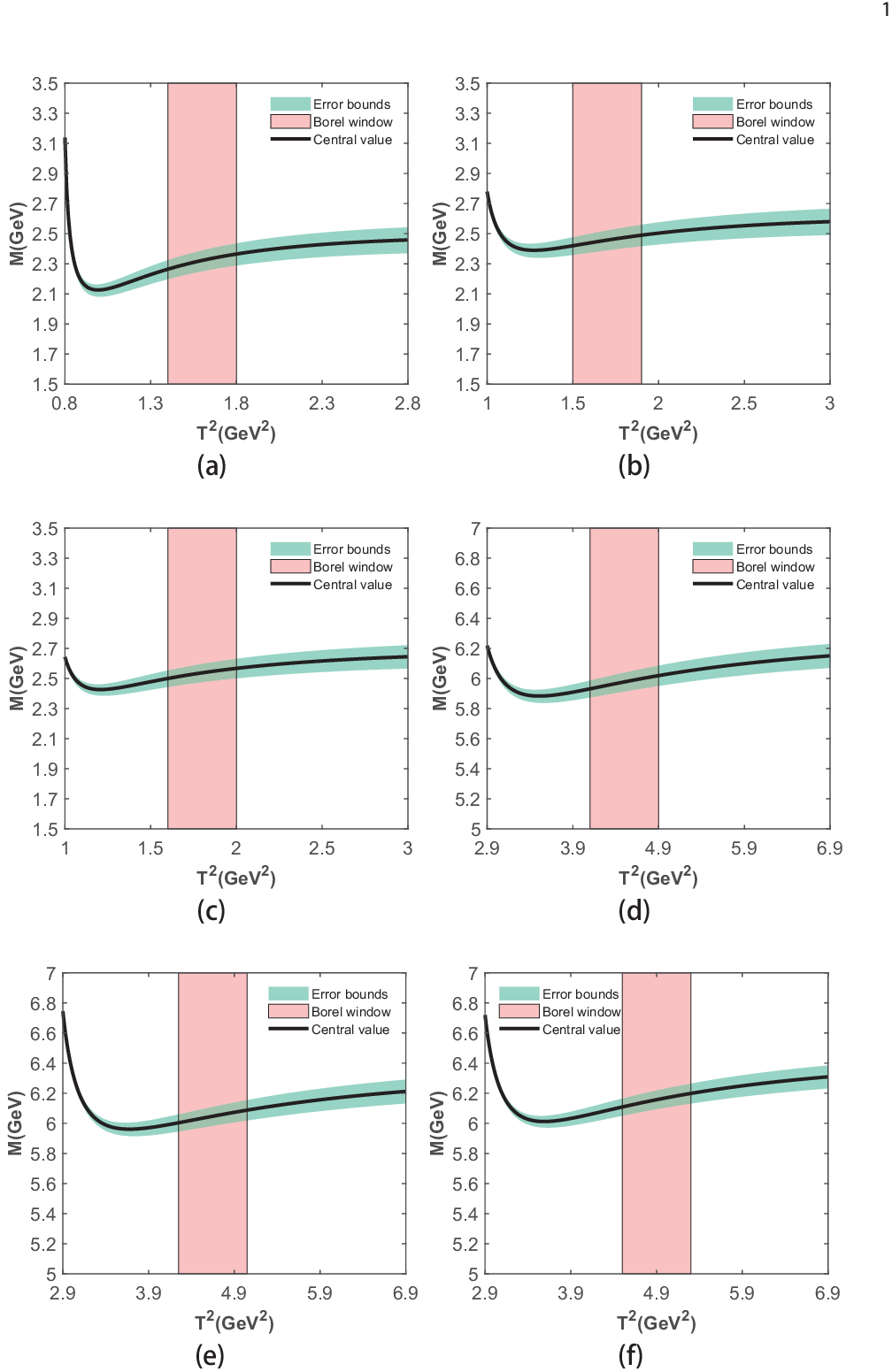}
	\caption{The masses of the tetraquark molecular states with variations of the Borel parameters $T^2$, where a, b, c, d, e and f denote $DK$, $D^*K$, $DK^*$, $BK$, $B^*K$ and $BK^*$ molecular states, respectively.}
	\label{composed picture of mass of current}
\end{figure}

\begin{figure}
	\centering
	\includegraphics[width=8.5cm, trim=0 0 20 30, clip]{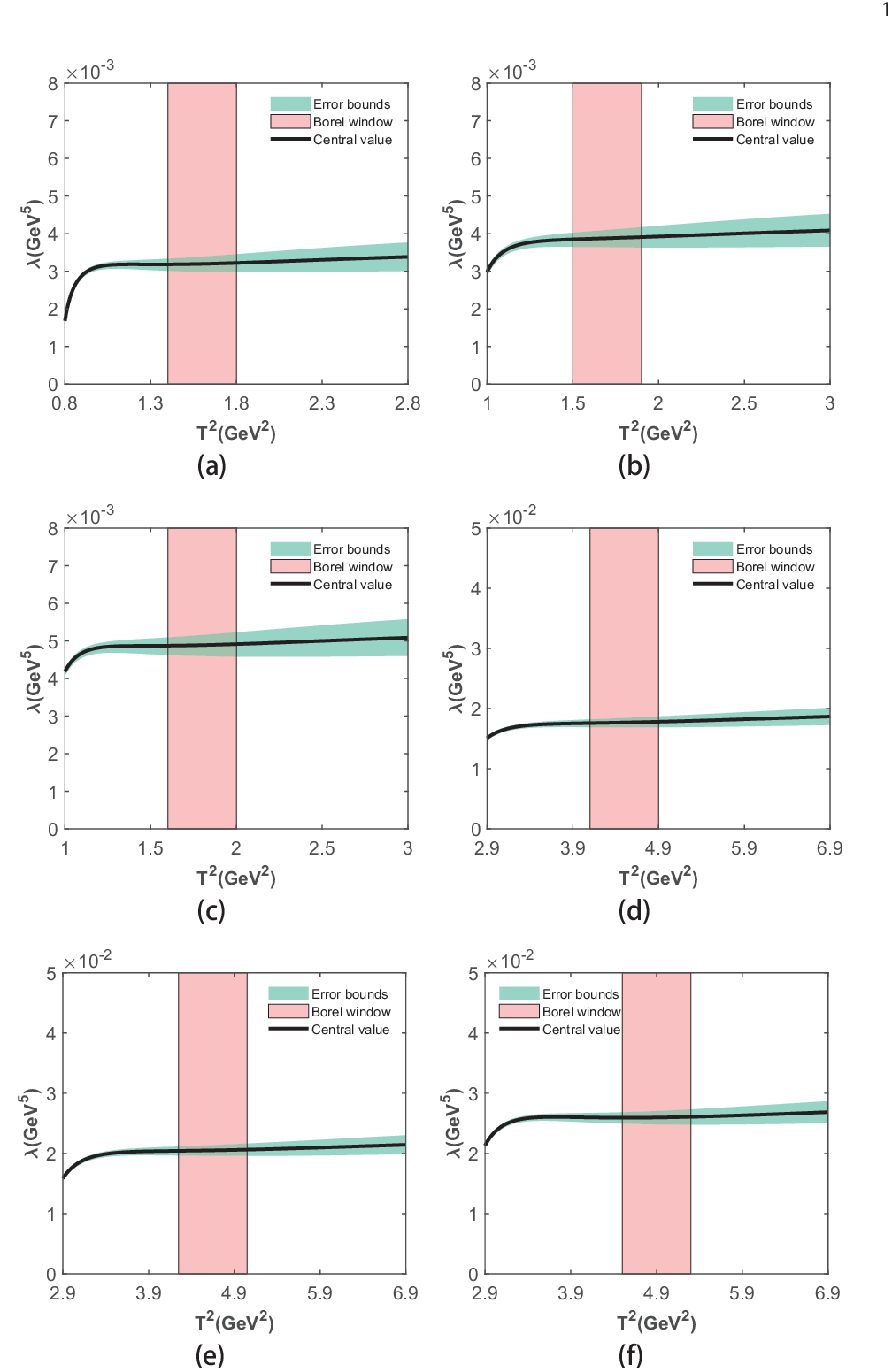}
	\caption{The pole residues of the tetraquark molecular states with variations of the Borel parameters $T^2$, where a, b, c, d, e and f denote $DK$, $D^*K$, $DK^*$, $BK$, $B^*K$ and $BK^*$ molecular states, respectively.}
	\label{composed pole residues of mass of current}
\end{figure}

\begin{figure}
	\centering
	\includegraphics[width=8cm, trim=30 20 30 25, clip]{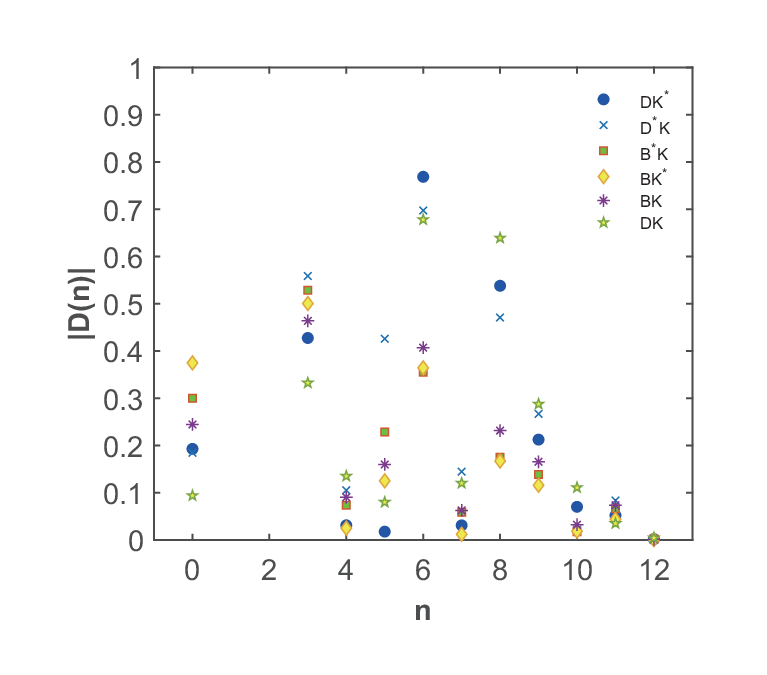}
	\caption{The absolute values of different dimensional contributions of tetraquark molecular states.}
	\label{condensation of 12 dimensions}
\end{figure}

The pole dominance requires that the pole contribution should be larger than 50$\%$, and the convergence of OPE requires the contribution of high dimensional condensate should be small. After repeated trial and contrast, the Borel windows are determined and are explicitly shown in Figs. \ref{composed picture of mass of current} and \ref{composed pole residues of mass of current}.
It can be seen from these figures that the results are weak dependent on Borel parameters and are stable in the Borel windows. The contributions originate from different vacuum condensate terms are shown in Fig. \ref{condensation of 12 dimensions}. From this figure, we notice that condensate terms with dimension n=3, 6 and 8 play important roles to the final results. It is also shown that contributions from condensate terms ($n>8$) become smaller with increase of $n$. Therefore, the convergence of OPE is satisfied. As for the pole contributions of different molecular states, we show them in Fig. \ref{pc of mass of six currents} in Appendix \ref{appendix A}. It is shown by these figures that the pole contributions are approximately $40\%$$-$$60\% $ in the Borel windows, and the central values exceed $50\%$. After all of the conditions of QCD sum rules are satisfied, we can extract reliable results which are listed in Table \ref{table for mass} together with the Borel parameter ($T^{2}$), threshold parameter ($s_{0}$), energy scale ($\mu$), pole contribution (Pole) and contribution of $|D(12)|$.

\begin{table*}
	\begin{ruledtabular}
				\centering
		\renewcommand{\arraystretch}{1.4}
		\caption{The masses, pole residues, Borel windows, continuum threshold parameters, energy scales, pole contributions and the contribution of $\left| {D(12)} \right|$ for molecular states.}
		\begin{tabular}{c c c c c c c c c}
			States & $J^P$ & M$ (\mathrm{GeV})$ & $\lambda  (\mathrm{GeV}{^5})$ & ${T^2}(\mathrm{GeV}{^2})$ & $\sqrt {{s_0}} (\mathrm{GeV})$ & $\mu (\mathrm{GeV})$ & Pole & $\left| {D(12)} \right|$\\
			\hline
			$DK$ & $0^ +$ & $2.322_{ - 0.072}^{ + 0.066}$ & $3.20_{ - 0.21}^{ + 0.19} \times {10^{ - 3}}$ & 1.40 $-$ 1.80 & $2.814 \pm 0.1$ & 1.2 & $(62 - 39)\% $ & $  \ll  1\% $\\
			$D^*K$ & $1^ +$ & $2.457_{ - 0.068}^{ + 0.064}$ & $3.88_{ - 0.24}^{ + 0.22} \times {10^{ - 3}}$ & 1.50 $-$ 1.90 & $2.960 \pm 0.1$ & 1.5 & $(62 - 40)\% $ & $  \ll  1\% $\\
			$DK^*$ & $1^ +$ & $2.538_{ - 0.062}^{ + 0.059}$ & $4.89_{ - 0.29}^{ + 0.27} \times {10^{ - 3}}$ & 1.60 $-$ 2.00 & $3.020 \pm 0.1$ & 1.6 & $(62 - 42)\% $ & $  \ll  1\% $\\
			$BK$ & $0^ +$ & $5.970_{ - 0.064}^{ + 0.061}$ & $1.77_{ - 0.08}^{ + 0.08} \times {10^{ - 2}}$ & 4.10 $-$ 4.90 & $6.650 \pm 0.1$ & 2.8 & $(59 - 41)\% $ & $  \ll  1\% $\\
			$B^*K$ & $1^ +$ & $6.050_{ - 0.064}^{ + 0.062}$ & $2.05_{ - 0.09}^{ + 0.09} \times {10^{ - 2}}$ & 4.25 $-$ 5.05 & $6.730 \pm 0.1$ & 2.9 & $(59 - 41)\% $ & $  \ll  1\% $\\
			$BK^*$ & $1^ +$ & $6.158_{ - 0.063}^{ + 0.061}$ & $2.60_{ - 0.10}^{ + 0.10} \times {10^{ - 2}}$ & 4.50 $-$ 5.30 & $6.848 \pm 0.1$ & 3.1 & $(59 - 42)\% $ & $  \ll  1\% $
		\end{tabular}
		\label{table for mass}
	\end{ruledtabular}
\end{table*}

From Tab. \ref{table for mass}, we can see that the experimental data of $D_{s0}^{* + }(2317)$, $D_{s1}^ + (2460)$, and $D_{s1}^ + (2536)$ are well reproduced by the theoretical results of $DK$, $D^*K$ and $DK^*$ tetraquark molecular states, respectively. These results support the explanation of $D_{s0}^{* + }(2317)$, $D_{s1}^ + (2460)$ and $D_{s1}^ + (2536)$ as $DK$, $D^*K$ and $DK^*$ tetraquark molecular states. As for their bottom analogs, two structures were observed at LHCb, that is, $B_{sJ}(6064)$ and $B_{sJ}(6114)$ if decays directly to the $B^{+}K^{-}$ final state, or $B_{sJ}(6109)$ and $B_{sJ}(6158)$ if instead proceeds through $B^{*+}K^{-}$ \cite{LHCb:2020pet}. The $B_{sJ}(6158)$ is quite close to the $BK^{*}$ threshold. In the present work, the predicted masses of $BK$, $B^*K$ and $BK^*$ are $5.970_{ - 0.064}^{ + 0.061}$ GeV, $6.050_{ - 0.064}^{ + 0.062}$ GeV and $6.158_{ - 0.063}^{ + 0.061}$ GeV, respectively, where theoretical value of $BK^*$ is consistent well with the measuring mass of $B_{sJ}(6158)$. In Ref. \cite{Kong:2021ohg}, the ${B_{sJ}}(6158)$ was also suggested to be a good candidate of $BK^{*}$ molecular state. Certainly, there is also the article that treated it as a conventional meson ${B_s}({1^3}{D_1})$  \cite{Hao:2025vmw}. The predicted masses for $BK$ and $B^{*}K$ are all higher than their thresholds, which indicates that $BK$ and $B^*K$ may be the resonance states rather than bound states.

On the other hand, although the experimental data of $D_{s0}^{*+ }(2317)$, $D_{s1}^+ (2460)$, $D_{s1}^+ (2536)$ and ${B_{sJ}}(6158)$ can be well reproduced by the molecular picture $DK$, $D^*K$, $DK^*$ and $BK^*$, explaining them as hadronic molecules still needs more theoretical analysis such as the partial widths and ratios. For example, the ratio of $\Gamma(D_{s1}(2460))\rightarrow D_{s0}(2317)\gamma$ and $\Gamma(D_{s1}(2460))\rightarrow D^{*}\pi^{0}$ was reported to be $\frac{\Gamma(D_{s1}(2460))\rightarrow D_{s0}(2317)\gamma}{\Gamma(D_{s1}(2460))\rightarrow D^{*}\pi^{0}}<0.58$ \cite{CLEO:2003ggt} and $\frac{\Gamma(D_{s1}(2460))\rightarrow D_{s0}(2317)\gamma}{\Gamma(D_{s1}(2460))\rightarrow D^{*}\pi^{0}}<0.22$ \cite{BaBar:2003cdx}. In addition, studying the production of these physical states from the nonleptonic $B$ decay process is also essential to recognize their inner structure. As for the two other predicted resonance states $BK$, $B^*K$, the information of decay width is also very attractive for us, which can help to discover these states in experiments in the future, and that is the work we need to carry out next.

\section{Conclusion}\label{sec4}

In summary, the masses and pole residues of $DK$, $D^*K$, $DK^*$ and their bottom analogs are analyzed by the QCD sum rules. To improve the reliability of the results, the vacuum condensations up to dimension 12 are considered, and the optimal energy scale is determined by using the energy scale formula $\mu  = \sqrt {M^2 - \mathbb{M}_{c/b}^2}  - k{\mathbb{M}_s}$. The experimental data of $D_{s0}^{* + }(2317)$, $D_{s1}^ + (2460)$, ${D}_{s1}^ + (2536)$ and $B_{sJ}(6158)$ are well reproduced by $DK$, $D^*K$, $DK^*$ and $BK^*$, respectively. The results imply that these physical states are good candidates for the hadronic molecular states. Besides, another two bottom-strange resonance states $BK$ and $B^*K$ are also predicted with their masses being $5.970_{ - 0.064}^{ + 0.061}$ GeV and $6.050_{ - 0.064}^{ + 0.062}$ GeV. Further theoretical analysis about the decay width of these charm/bottom strange hadronic molecules is very helpful to confirm the inner structure of these physical states. The predicted pole residues in the present work are important input parameters in three-point QCD sum rules which can be used to study the decay properties of these hadronic molecules.

\section*{Acknowledgements}

This project is supported by National Natural Science Foundation, Grant Number 12175068 and Natural Science Foundation of HeBei Province, Grant Number A2024502002.

\begin{widetext}
\appendix
\section{The graphs of pole contribution}\label{appendix A}
\begin{figure*}[h!]
	\centering
	\includegraphics[width=19cm, trim=0 0 20 30, clip]{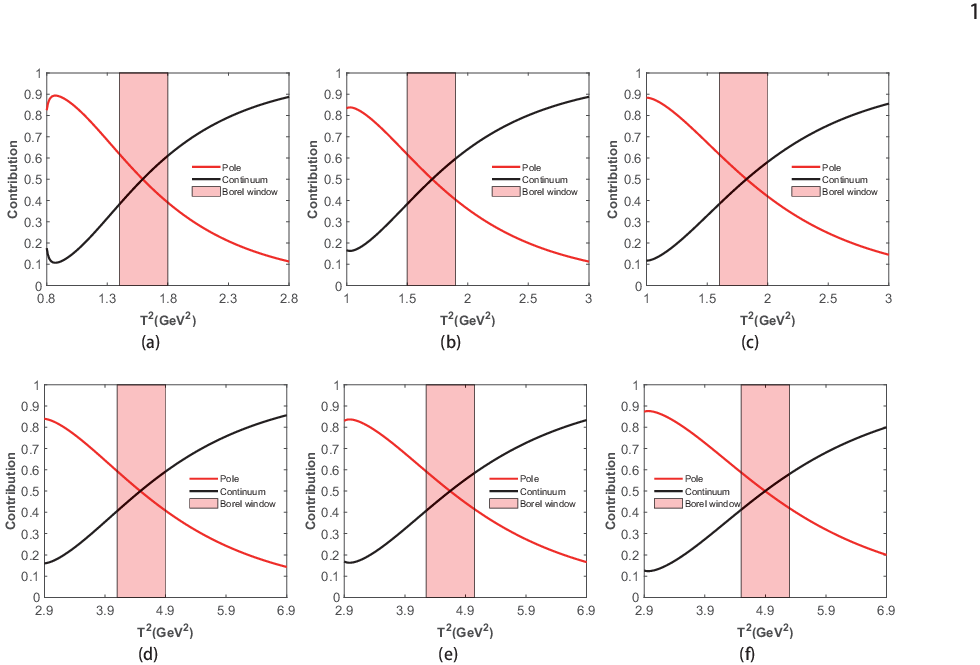}
	\caption{The pole and continuum contributions with variations of the Borel parameters, where a, b, c, d, e, and f denote $DK$, $D^*K$, $DK^*$, $BK$, $B^*K$ and $BK^*$ molecular states, respectively.}
	\label{pc of mass of six currents}
\end{figure*}
\FloatBarrier

\section{The expression of QCD spectral density}\label{appendix B}
	Due to the complexity of spectral densities, we only show the spectral densities of $DK$ molecular state as follows,
	\begin{align}
		\rho _{DK}^0 = \int_{\frac{{m_c^2}}{s}}^1 {dx} \frac{1}{{8192{\pi ^6}{x^3}}}{(x - 1)^4}{(sx - m_c^2)^3}(3sx - m_c^2)
	\end{align}
\begin{align}
	\notag
	\rho _{DK}^3 =& \int_{\frac{{m_c^2}}{s}}^1 {dx} \frac{{ - {{(x - 1)}^2}(sx - m_c^2)}}{{3072{\pi ^4}{x^2}}}\left( {2m_c^4{m_s}\left\langle {\bar qq} \right\rangle (x - 1) + m_c^2{m_s}x[36\left\langle {\bar ss} \right\rangle  - 7\left\langle {\bar qq} \right\rangle s(x - 1)]} \right)\\
	 &+\int_{\frac{{m_c^2}}{s}}^1 {dx} \frac{{ - {{(x - 1)}^2}(sx - m_c^2)}}{{3072{\pi ^4}{x^2}}}\left( {36{m_c}\left\langle {\bar qq} \right\rangle sx + {m_s}s{x^2}[5\left\langle {\bar qq} \right\rangle s(x - 1) - 72\left\langle {\bar ss} \right\rangle ] - 36m_c^3\left\langle {\bar qq} \right\rangle } \right)
\end{align}
\begin{align}
	\rho _{DK}^4 = \int_{\frac{{m_c^2}}{s}}^1 {dx} \frac{{\left\langle {g_s^2GG} \right\rangle {{(x - 1)}^2}}}{{49152{\pi ^6}{x^3}}}[m_c^4(26{x^2} + 8x + 2) - 3sx{(5{m_c}x + {m_c})^2} + 24{s^2}{x^3}(2x + 1)]
\end{align}
\begin{align}
	\notag
	\rho _{DK}^5 =& \int_{\frac{{m_c^2}}{s}}^1 {dx} \frac{{(1 - x)}}{{2048{\pi ^4}{x^2}}}\left( {3m_c^4{m_s}\left\langle {\bar qgs\sigma Gq} \right\rangle (x - 1)x + m_c^2{m_s}{x^2}[16\left\langle {\bar sgs\sigma Gs} \right\rangle  - 9\left\langle {\bar qgs\sigma Gq} \right\rangle s(x - 1)]} \right)\\
	\notag
	&+\int_{\frac{{m_c^2}}{s}}^1 {dx} \frac{{(1 - x)}}{{2048{\pi ^4}{x^2}}}[24{m_c}\left\langle {\bar qgs\sigma Gq} \right\rangle sx(2x - 1) + 24m_c^3\left\langle {\bar qgs\sigma Gq} \right\rangle (1 - 2x)]\\
	&+\int_{\frac{{m_c^2}}{s}}^1 {dx} \frac{{(1 - x)x}}{{1024{\pi ^4}}}3{m_s}s[\left\langle {\bar qgs\sigma Gq} \right\rangle s(x - 1) - 4\left\langle {\bar sgs\sigma Gs} \right\rangle ]
	\end{align}
\begin{align}
	\notag
	\rho _{DK}^6 =& \int_{\frac{{m_c^2}}{s}}^1 {dx} \frac{1}{{1728{\pi ^4}x}}[2m_c^2(x - 1)x - 3s(x - 1){x^2}][g_s^2(2{\left\langle {\bar qq} \right\rangle ^2} + {\left\langle {\bar ss} \right\rangle ^2}) + 108{\pi ^2}\left\langle {\bar qq} \right\rangle \left\langle {\bar ss} \right\rangle ]\\
	&+\int_{\frac{{m_c^2}}{s}}^1 {dx} \frac{1}{{1728{\pi ^4}x}}\left( { - 27{\pi ^2}m_c^3{m_s}{{\left\langle {\bar qq} \right\rangle }^2}(x - 1) + 27{\pi ^2}{m_c}{m_s}\left\langle {\bar qq} \right\rangle x[\left\langle {\bar qq} \right\rangle s(x - 1) - 2\left\langle {\bar ss} \right\rangle ]} \right)
\end{align}
\begin{align}
	\notag
	\rho _{DK}^7 =& \int_{\frac{{m_c^2}}{s}}^1 {dx} \frac{{\left\langle {g_s^2GG} \right\rangle m_c^2(x - 1)\delta (s - \frac{{m_c^2}}{x})}}{{36864{\pi ^4}{x^3}}}m_c^2{m_s}(x - 1)[\left\langle {\bar qq} \right\rangle (x - 1) - \frac{{6\left\langle {\bar ss} \right\rangle }}{T^2}]\\
	\notag
	&+\int_{\frac{{m_c^2}}{s}}^1 {dx} \frac{{\left\langle {g_s^2GG} \right\rangle m_c^2(x - 1)\delta (s - \frac{{m_c^2}}{x})}}{{36864{\pi ^4}{x^3}}}[12{m_c}\left\langle {\bar qq} \right\rangle (x - 1) + 6{m_s}\left\langle {\bar ss} \right\rangle (1 - 7x)x]\\
	\notag
	&-\int_{\frac{{m_c^2}}{s}}^1 {dx} \frac{{\left\langle {g_s^2GG} \right\rangle }}{{36864{\pi ^4}{x^2}}}[2m_c^2{m_s}\left\langle {\bar qq} \right\rangle (5{x^3} - 12{x^2} + 6x + 1) + 12{m_c}\left\langle {\bar qq} \right\rangle (10{x^2} - 6x + 3)]\\
	&+\int_{\frac{{m_c^2}}{s}}^1 {dx} \frac{{\left\langle {g_s^2GG} \right\rangle }}{{36864{\pi ^4}}}9{m_s}(x - 1)[\left\langle {\bar qq} \right\rangle s(2x - 3) - 8\left\langle {\bar ss} \right\rangle ]
	\end{align}
\begin{align}
	\notag
	\rho _{DK}^8 =& \int_{\frac{{m_c^2}}{s}}^1 {dx} \frac{x}{{8192{\pi ^6}x}}[ - {\left\langle {g_s^2GG} \right\rangle ^2}(x - 1) - 256{\pi ^4}x(\left\langle {\bar qgs\sigma Gq} \right\rangle \left\langle {\bar ss} \right\rangle  + \left\langle {\bar qq} \right\rangle \left\langle {\bar sgs\sigma Gs} \right\rangle )]\\
	\notag
	&+\int_{\frac{{m_c^2}}{s}}^1 {dx} \frac{1}{{8192{\pi ^6}x}}64{\pi ^4}{m_c}{m_s}\left\langle {\bar qgs\sigma Gq} \right\rangle \left\langle {\bar qq} \right\rangle (2x - 1)\\
	\notag
	&-\int_{\frac{{m_c^2}}{s}}^1 {dx} \frac{{{m_c}\delta (s - \frac{{m_c^2}}{x})}}{{98304{\pi ^6}T^2{x^3}}}{\left\langle {g_s^2GG} \right\rangle ^2}{m_c}(x - 1)[m_c^2(x - 1) + T^2x(7x - 1)]\\
	\notag
	&-\int_{\frac{{m_c^2}}{s}}^1 {dx} \frac{{{m_c}\delta (s - \frac{{m_c^2}}{x})}}{{98304{\pi ^6}x}}1536{\pi ^4}({m_c}x\left\langle {\bar qgs\sigma Gq} \right\rangle \left\langle {\bar ss} \right\rangle  + {m_c}x\left\langle {\bar qq} \right\rangle \left\langle {\bar sgs\sigma Gs} \right\rangle  + {m_s}\left\langle {\bar qgs\sigma Gq} \right\rangle \left\langle {\bar ss} \right\rangle )\\
	&+\frac{{{m_c}{m_s}\delta (s - m_c^2)(3\left\langle {\bar qgs\sigma Gq} \right\rangle \left\langle {\bar ss} \right\rangle  + 2\left\langle {\bar qq} \right\rangle \left\langle {\bar sgs\sigma Gs} \right\rangle )}}{{384{\pi ^2}}}
\end{align}
\begin{align}
	\notag
   \rho _{DK}^9 =& \int_{\frac{{m_c^2}}{s}}^1 {dx} \frac{{\left\langle {g_s^2GG} \right\rangle \delta (s - \frac{{m_c^2}}{x})}}{{147456{\pi ^4}{T^4}{x^3}}}\left( {m_c^4{m_s}(x - 1)[3\left\langle {\bar qgs\sigma Gq} \right\rangle T^2(x - 1) - 8\left\langle {\bar sgs\sigma Gs} \right\rangle ] + 24m_c^3\left\langle {\bar qgs\sigma Gq} \right\rangle T^2(x - 1)} \right)\\
   \notag
   &+\int_{\frac{{m_c^2}}{s}}^1 {dx} \frac{{\left\langle {g_s^2GG} \right\rangle \delta (s - \frac{{m_c^2}}{x})}}{{49152{\pi ^4}T^2{x^2}}}m_c^2{m_s}[\left\langle {\bar qgs\sigma Gq} \right\rangle T^2(7{x^2} - 8x + 1) - 8\left\langle {\bar sgs\sigma Gs} \right\rangle x]\\
   \notag
   &-\int_{\frac{{m_c^2}}{s}}^1 {dx} \frac{{\left\langle {g_s^2GG} \right\rangle \delta (s - \frac{{m_c^2}}{x})}}{{6144{\pi ^4}{x^2}}}(3{m_c}\left\langle {\bar qgs\sigma Gq} \right\rangle (3x - 1) + {m_s}\left\langle {\bar sgs\sigma Gs} \right\rangle {x^2})\\
   \notag
   &+\int_{\frac{{m_c^2}}{s}}^1 {dx} \frac{1}{{5184{\pi ^2}}}\delta (s - \frac{{m_c^2}}{x})g_s^2m_c^2{m_s}{\left\langle {\bar qq} \right\rangle ^3}\\
   \notag
   &+\int_{\frac{{m_c^2}}{s}}^1 {dx} [\frac{{\left\langle {g_s^2GG} \right\rangle {m_s}\left\langle {\bar qgs\sigma Gq} \right\rangle (x - 1)}}{{4096{\pi ^4}}} + \frac{{g_s^2{m_s}{{\left\langle {\bar qq} \right\rangle }^3}x}}{{2592{\pi ^2}}}]\\
   \notag
   &+\delta (s - m_c^2)(\frac{{\left\langle {g_s^2GG} \right\rangle {m_c}\left\langle {\bar qgs\sigma Gq} \right\rangle }}{{2048{\pi ^4}}} - \frac{{{m_c}{{\left\langle {\bar qq} \right\rangle }^2}\left\langle {\bar ss} \right\rangle }}{{12}} + \frac{{g_s^2\left\langle {\bar qq} \right\rangle m_c^2{m_s}\left\langle {\bar qq} \right\rangle \left\langle {\bar ss} \right\rangle }}{{1296{\pi ^2}T^2}})\\
   &+\delta (s - m_c^2)\frac{{g_s^2\left\langle {\bar qq} \right\rangle [ - {m_c}({{\left\langle {\bar qq} \right\rangle }^2} + {{\left\langle {\bar ss} \right\rangle }^2}) + {m_s}\left\langle {\bar qq} \right\rangle \left\langle {\bar ss} \right\rangle ]}}{{1296{\pi ^2}}}
\end{align}
\begin{align}
	\notag
	\rho _{DK}^{10} =& \int_{\frac{{m_c^2}}{s}}^1 {dx} \frac{{\delta (s - \frac{{m_c^2}}{x})\left\langle {g_s^2GG} \right\rangle g_s^2}}{{124416{\pi ^4}{T^4}{x^3}}}[m_c^4(x - 1)(2{\left\langle {\bar qq} \right\rangle ^2} + {\left\langle {\bar ss} \right\rangle ^2}) + 3m_c^2T^2{x^2}({\left\langle {\bar qq} \right\rangle ^2} + {\left\langle {\bar ss} \right\rangle ^2}) + 3{T^4}{x^3}({\left\langle {\bar qq} \right\rangle ^2} + {\left\langle {\bar ss} \right\rangle ^2})]\\
	\notag
	&+\int_{\frac{{m_c^2}}{s}}^1 {dx} \frac{{\delta (s - \frac{{m_c^2}}{x}){\pi ^2}\left\langle {g_s^2GG} \right\rangle \left\langle {\bar qq} \right\rangle }}{{4608{\pi ^4}{T^4}{x^3}}}\left( {4m_c^4\left\langle {\bar ss} \right\rangle (x - 1) + m_c^3{m_s}[2\left\langle {\bar ss} \right\rangle  - \left\langle {\bar qq} \right\rangle T^2(x - 1)] + 12m_c^2\left\langle {\bar ss} \right\rangle T^2{x^2}} \right)\\
	\notag
	&+\int_{\frac{{m_c^2}}{s}}^1 {dx} \frac{{\delta (s - \frac{{m_c^2}}{x})}}{{1536{\pi ^4}T^2{x^2}}}{\pi ^2}\left\langle {g_s^2GG} \right\rangle \left\langle {\bar qq} \right\rangle \left( {{m_c}{m_s}[\left\langle {\bar qq} \right\rangle T^2(x - 1) - 2\left\langle {\bar ss} \right\rangle ] + 4\left\langle {\bar ss} \right\rangle T^2{x^2}} \right)\\
	\notag
	&+\int_{\frac{{m_c^2}}{s}}^1 {dx} \frac{{\delta (s - \frac{{m_c^2}}{x})}}{{512{\pi ^2}{x^2}}}{m_c}{m_s}{\left\langle {\bar qgs\sigma Gq} \right\rangle ^2}\\
	\notag
	&+\frac{{\delta (s - m_c^2)}}{{27648{\pi ^4}T^2}}(m_c^2 + T^2)[\left\langle {g_s^2GG} \right\rangle \left\langle {\bar qq} \right\rangle (g_s^2\left\langle {\bar qq} \right\rangle  + 36{\pi ^2}\left\langle {\bar ss} \right\rangle ) + 162{\pi ^2}\left\langle {\bar qgs\sigma Gq} \right\rangle \left\langle {\bar sgs\sigma Gs} \right\rangle ]\\
	\notag
	&-\frac{{\delta (s - m_c^2)}}{{4608{\pi ^2}{T^4}}}m_c^3{m_s}(\left\langle {g_s^2GG} \right\rangle \left\langle {\bar qq} \right\rangle \left\langle {\bar ss} \right\rangle  + 6\left\langle {\bar qgs\sigma Gq} \right\rangle \left\langle {\bar sgs\sigma Gs} \right\rangle )\\
	&-\frac{{\delta (s - m_c^2)}}{{9216{\pi ^2}T^2}}{m_c}{m_s}(2\left\langle {g_s^2GG} \right\rangle {\left\langle {\bar qq} \right\rangle ^2}T^2 + 9{\left\langle {\bar qgs\sigma Gq} \right\rangle ^2}T^2 - 24\left\langle {\bar qgs\sigma Gq} \right\rangle \left\langle {\bar sgs\sigma Gs} \right\rangle )
\end{align}
    \begin{align}
    	\notag
    	\rho _{DK}^{11} =& \int_{\frac{{m_c^2}}{s}}^1 {dx} \frac{{{{\left\langle {g_s^2GG} \right\rangle }^2}\left\langle {\bar qq} \right\rangle \delta (s - \frac{{m_c^2}}{x})}}{{884736{\pi ^4}{T^4}{x^3}}}(m_c^4{m_s}(x - 1) + 28m_c^3 + 3m_c^2{m_s}T^2{x^2} - 84{m_c}T^2x + 3{m_s}{T^4}{x^3})\\
    	\notag
    	&+\frac{{\delta (s - m_c^2){m_c}\left\langle {\bar qq} \right\rangle }}{{48{T^4}}}[m_c^2(2\left\langle {\bar qgs\sigma Gq} \right\rangle \left\langle {\bar ss} \right\rangle  + \left\langle {\bar qq} \right\rangle \left\langle {\bar sgs\sigma Gs} \right\rangle ) - 2\left\langle {\bar qgs\sigma Gq} \right\rangle \left\langle {\bar ss} \right\rangle T^2]\\
    	\notag
    	&-\frac{{\delta (s - m_c^2)g_s^2}}{{15552{\pi ^2}{T^6}}}[2m_c^4{m_s}{\left\langle {\bar qq} \right\rangle ^2}\left\langle {\bar sgs\sigma Gs} \right\rangle  - 3m_c^3\left\langle {\bar qgs\sigma Gq} \right\rangle T^2({\left\langle {\bar qq} \right\rangle ^2} + {\left\langle {\bar ss} \right\rangle ^2})]\\
    	&-\frac{{\delta (s - m_c^2)g_s^2}}{{20736{\pi ^2}T^2}}(m_c^2{m_s} + 16{m_c} + {m_s}T^2)\left\langle {\bar qgs\sigma Gq} \right\rangle {\left\langle {\bar qq} \right\rangle ^2} - \frac{{\delta (s - m_c^2)m_c^3\left\langle {\bar qq} \right\rangle }}{{73728{\pi ^4}{T^4}}}{\left\langle {g_s^2GG} \right\rangle ^2}
    \end{align}
   \begin{align}
   	\notag
   	\rho _{DK}^{12} =&  - \int_{\frac{{m_c^2}}{s}}^1 {dx} \frac{{\left\langle {g_s^2GG} \right\rangle {m_c}\delta (s - \frac{{m_c^2}}{x})}}{{9216{\pi ^2}{T^6}{x^3}}}[(2{m_c}T^2x - 2m_c^3)(\left\langle {\bar qgs\sigma Gq} \right\rangle \left\langle {\bar ss} \right\rangle  + \left\langle {\bar qq} \right\rangle \left\langle {\bar sgs\sigma Gs} \right\rangle )]\\
   	\notag
   	&-\int_{\frac{{m_c^2}}{s}}^1 {dx} \frac{{\left\langle {g_s^2GG} \right\rangle {m_c}\delta (s - \frac{{m_c^2}}{x})}}{{9216{\pi ^2}{T^6}{x^3}}}[(m_c^2 + 1){m_s}T^2\left\langle {\bar qgs\sigma Gq} \right\rangle \left\langle {\bar qq} \right\rangle ]\\
   	\notag
   	&+\frac{{\delta (s - m_c^2)\left\langle {g_s^2GG} \right\rangle m_c^4}}{{27648{\pi ^2}{T^8}}}[m_c^3{m_s}\left\langle {\bar qq} \right\rangle \left\langle {\bar sgs\sigma Gs} \right\rangle  - 3T^2(\left\langle {\bar qgs\sigma Gq} \right\rangle \left\langle {\bar ss} \right\rangle  + \left\langle {\bar qq} \right\rangle \left\langle {\bar sgs\sigma Gs} \right\rangle )]\\
   	\notag
   	&+\frac{{\delta (s - m_c^2)\left\langle {g_s^2GG} \right\rangle m_c^3{m_s}}}{{55296{\pi ^2}{T^6}}}(3\left\langle {\bar qgs\sigma Gq} \right\rangle \left\langle {\bar qq} \right\rangle T^2 - 6\left\langle {\bar qgs\sigma Gq} \right\rangle \left\langle {\bar ss} \right\rangle  - 8\left\langle {\bar qq} \right\rangle \left\langle {\bar sgs\sigma Gs} \right\rangle )\\
   	\notag
   	&-\frac{{\delta (s - m_c^2)\left\langle {g_s^2GG} \right\rangle m_c^2}}{{1536{\pi ^2}{T^4}}}(\left\langle {\bar qgs\sigma Gq} \right\rangle \left\langle {\bar ss} \right\rangle  + \left\langle {\bar qq} \right\rangle \left\langle {\bar sgs\sigma Gs} \right\rangle )\\
   	\notag
   	&+\frac{{\delta (s - m_c^2)\left\langle {g_s^2GG} \right\rangle {m_c}{m_s}}}{{18432{\pi ^2}{T^4}}}( - \left\langle {\bar qgs\sigma Gq} \right\rangle \left\langle {\bar qq} \right\rangle T^2 + 6\left\langle {\bar qgs\sigma Gq} \right\rangle \left\langle {\bar ss} \right\rangle  + 4\left\langle {\bar qq} \right\rangle \left\langle {\bar sgs\sigma Gs} \right\rangle )\\
   	&-\frac{{\delta (s - m_c^2)g_s^2m_c^3}}{{1944{T^8}}}{\left\langle {\bar qq} \right\rangle ^3}\left\langle {\bar ss} \right\rangle [m_c^2{m_s} - 2({m_c} + {m_s}){T^2}]
   \end{align}
\end{widetext}
	
\end{document}